\documentclass{aastex631}

\shorttitle{Interactive GUI tool for galaxy spectra visualization}
\shortauthors{Leung et al.}
\graphicspath{{./}{figures/}}

\begin{document}

\title{Introducing a Real-time Interactive GUI Tool for Visualization of Galaxy Spectra}

\correspondingauthor{Ho-Hin Leung}
\email{hhl1@st-andrews.ac.uk}

\author{Ho-Hin Leung}

\author[0000-0002-8956-7024]{Vivienne Wild}
\affiliation{SUPA\footnote{Scottish Universities Physics Alliance}, School of Physics \& Astronomy, 
University of St Andrews, \\
North Haugh, St Andrews, Fife KY16 9SS, UK}

\author{Adam Carnall}
\affiliation{SUPA, Institute for Astronomy, University of Edinburgh, \\
Royal Observatory, 
Edinburgh, EH9 3HJ, UK}

\author[0000-0002-5897-695X]{Michail Papathomas}
\affiliation{School of Mathematics and Statistics, University of St Andrews, \\
North Haugh, St Andrews, 
Fife, KY16 9SS, UK}



\begin{abstract}
To aid the understanding of the non-linear relationship between galaxy properties and predicted spectral energy distributions (SED), we present a new interactive graphical user interface (GUI) tool \textit{pipes\_vis} based on \textsc{Bagpipes} \citep{bagpipes2018,bagpipes2019}. It allows for real-time manipulation of a model galaxy's star formation history, dust and other relevant properties through sliders and text boxes, with each change's effect on the predicted SED reflected instantaneously. We hope the tool will assist in building intuition about what affects the SED of galaxies, potentially helping to speed up fitting stages such as prior construction, and aid in undergraduate and graduate teaching. \textit{pipes\_vis} is available online\footnote[2]{\textit{pipes\_vis} is maintained and documented online at \url{https://github.com/HinLeung622/pipes_vis}, or version 0.4.1 is archived in Zenodo \citep{pipes_vis} and also available for installation through \texttt{pip install pipes\_vis}}.
\end{abstract}

\keywords{Galaxy evolution, Spectral energy distribution, Astronomy data visualization}


\section{Introduction}
The study of galaxy evolution, aimed at understanding the processes that drive changes in galaxies, often requires the recovery of galaxy properties such as the total stellar mass, star-formation history (SFH), chemical evolution and dust properties, that are not directly measurable from the observed spectral energy distributions (SEDs). This is often done in a fitting process that relies on stellar population synthesis (SPS) models, which predict a galaxy's stellar-contributed SED when given its underlying properties and assumed initial mass function, isochrone tables, spectral libraries, etc. Non-stellar components, such as nebular and dust emission are also included via energy balance or ionisation balance constraints \citep[for a comprehensive review, see][]{conroy2013}. This fitting process is complex and highly multi-dimensional, and can result in complex degeneracies between parameters \citep{ocvirk2006,pacifici2015,carnall2019,leja2019}. In fitting contexts, the likelihood is also typically multi-modal (multi-peaked, each with similar likelihood levels), leading to multi-modal posteriors in Bayesian methods of fitting. Hence, a tool that quickly visualizes the effects on the predicted SED from changing each individual galaxy property will supply intuition for this non-linear transformation, possibly allowing for more efficient prior construction during fitting, and limit the amount of trial-and-error required before producing robust results. 

\textsc{Sengi} \citep{sengi} is an example of a similar visualizer, which works as a web-based online tool displaying changes in the SED in real-time. \textsc{Sengi} focuses on providing modelled SEDs from multiple SPS codes (FSPS \citep{FSPS}, BC03 \citep{BC03}, etc.), but is only able to model the stellar continuum SEDs of simple stellar populations (SSPs, coeval stellar populations with the same age, metallicity and chemical composition). Approaching from the fitting point of view with our tool, we aim to provide SED predictions from complex SFHs forms, while accounting for non-stellar elements including dust attenuation, emission and nebular line emission that are all crucially important for fitting, and allow for real-time adjustment for all parameters in these models. However, we rely on stellar continuum predictions from one SPS code alone (BC03), and the range of dust emission and nebular emission options are arguably limited compared to the true diversity of the galaxy population.

\section{A brief description of the visualizer tool}
\begin{figure}
    \plotone{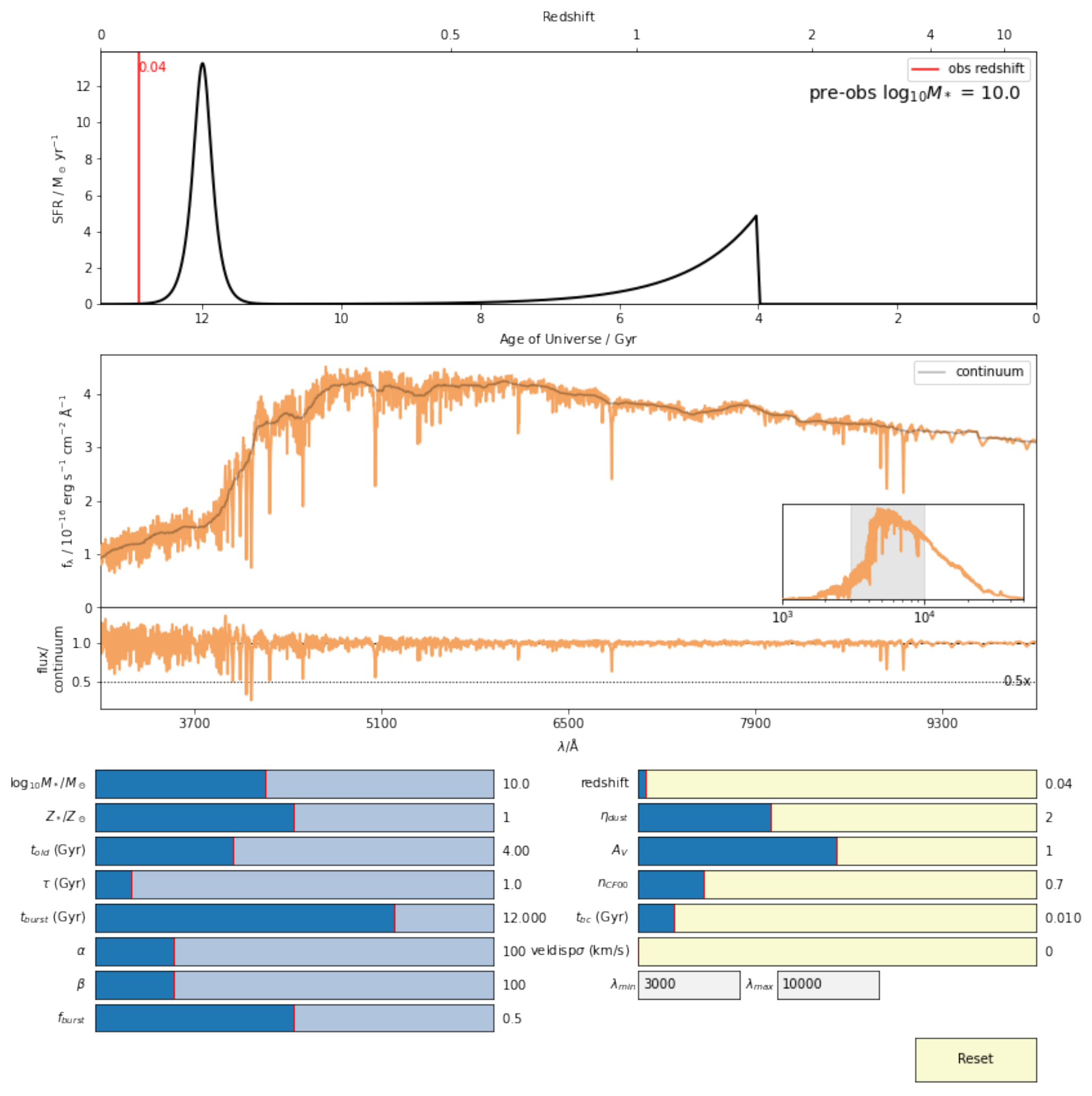}
    \caption{An example of the main GUI of \textit{pipes\_vis}. The top panel plots the user-defined SFH of the galaxy, from the beginning of the Universe to $z=0$. The total stellar mass formed prior to the time of observation (red vertical line) is given in the top right. The central panel plots the model-predicted spectrum, with wavelength range as specified by the text boxes underneath. This panel also corresponds to the shaded region in the full spectrum subplot, which is plotted in log wavelength scale. The bottom panel plots the central panel's flux divided by its continuum against wavelength, aimed to highlight absorption and emission bands. These panels are updated whenever any slider or text box on the lower half is changed. For this example, a 2-component post-starburst SFH functional form is used \citep{wild2020}, which consists of the 8 parameters on the left column, along with a CF00 \citep{CF00} dust law. The slider symbols have their usual meanings, while $\eta_{dust}$ means the multiplicative factor on $A_V$ for stars in birth clouds, $n_{CF00}$ the power-law slope of CF00 dust attenuation and $t_{bc}$ the maximum age of stellar birth clouds. If a SFH function with more parameters and/or multiple SFH components were to be used, the left column would rearrange into multiple columns to accommodate this, and use different slider colours to differentiate between SFH components. \label{fig:GUI}}
\end{figure}

Figure \ref{fig:GUI} shows an example of the main GUI interface when using \textit{pipes\_vis}, which is maintained and documented online\footnote{\url{https://github.com/HinLeung622/pipes_vis}}, or version 0.4.1\footnote{Also available through \texttt{pip install pipes\_vis}} is archived in Zenodo \citep{pipes_vis}. The top panel plots the SFH, from the beginning of the Universe to $z=0$. The vertical red line shows the time and redshift of observation, which moves horizontally with the redshift slider controlled by the user. Only light from stellar populations formed rightwards of the line contributes to the predicted SED. The central panel shows the predicted SED at the wavelength limits controlled by the bottom right text boxes, also indicated with the shaded region within the full SED subplot. The bottom panel shows the fraction flux/continuum, aimed to highlight absorption and emission lines. The continuum is calculated via a rolling median, with a width of 150\AA. 

Written in \textsc{Python}, the visualizer is in practice a wrapper of the \textsc{Python} fully-Bayesian SED fitting code \textsc{Bagpipes} \citep{bagpipes2018,bagpipes2019}, relying on \textsc{Bagpipes} to re-calculate predicted SEDs mapped onto a defined wavelength array upon each change in the sliders. The main functionalities are as follows:
\begin{itemize}
    \item The GUI mentioned above.
    \item Returning the GUI plots as a static image without sliders,  for users who might want it for fixed display or further manipulation.
    \item Plotting all predicted SEDs when sweeping across a defined range of a single parameter. This displays the changes each parameter can bring about, while keeping every other parameters fixed.
\end{itemize}

Since it is built on top of \textsc{Bagpipes}, \textit{pipes\_vis} is designed to have similar model creation conventions as the underlying code. However, an important caveat regards parameters that refer to particular points in time: due to the time-invariant nature of the inputted SFH, \textit{pipes\_vis} requires inputs to be in age of the Universe (Gyr), instead of the more commonly used lookback time (Gyr). This change does not affect temporal parameters that refer to durations (e.g. decay timescale of the exponential function). Like \textsc{Bagpipes}, \textit{pipes\_vis} supports the stacking of multiple SFH components of both different and the same functional forms, by generating individual sets of sliders for each component used. In line with \textsc{Bagpipes}, \textit{pipes\_vis} assumes a $\Lambda$CDM cosmology with $\Omega_M=0.3$, $\Omega_\Lambda=0.7$ and $h=0.7$. 

Currently, \textit{pipes\_vis} only predicts spectroscopic SEDs, but it would be possible to add predictions for photometric bands with user-specified filter curves, and Lick indices, in a similar setup in the future should there be interest from the community.


\bibliography{sample631}{}
\bibliographystyle{aasjournal}



\end{document}